# Composition dependence of magnetocaloric effect in $Sm_{1-x}Sr_xMnO_3$ (x = 0.3-0.5)

A.Rebello[1] and R. Mahendiran[1,2]

[1]Department of Physics, Faculty of Science, National University of Singapore, 2 Science Drive 3, Singapore -117542

[2]NUS Nanoscience &Nanotechnology Initiative (NUSNNI), Faculty of Science, National University of Singapore, 2 Science Drive 3, Singapore -117542

**Abstract**

We investigated magnetic and magnetocaloric properties in $Sm_{1-x}Sr_xMnO_3$ (x = 0.30-0.5). We report a magnetic field driven first-order metamagnetic transition in the paramagnetic state in x = 0.4 and 0.5 and a second-order transition in x = 0.3. The highest magnetic entropy ($-\Delta S_m$ = 6.2 J/kgK for $\Delta H$ = 5 T at $T$ = 125 K) that occurs in x = 0.4 is associated with the metamagnetic transition resulting from the field-induced growth and coalescence of ferromagnetic nano clusters preexisting in the paramagnetic state. Our results suggest that manganites with intrinsic nanoscale phase separation can be exploited for magnetic refrigeration.

When a material is magnetized by the application of a magnetic field, the entropy associated with the magnetic degree of freedom, $S_m$, decreases. Under adiabatic conditions, the entropy change, $\Delta S_m$, must be compensated for by an equal but opposite change in the entropy associated with the lattice, $\Delta S_l$, resulting in a change in temperature of the material, $\Delta T$, referred to as the magnetocaloric effect. Magnetic refrigeration based on the magnetocaloric effect is a promising alternative technology to the conventional vapor-compression refrigeration because of its higher energy efficiency and environmental friendliness.[1] The change in magnetic entropy ($\Delta S_M = \int_0^H \left(\frac{\partial M}{\partial T}\right)_H dH$) and hence MCE is expected to be maximum at the paramagnetic to ferromagnetic transition temperature ($T_C$). Majority of materials show a second–order paramagnetic to ferromagnetic transition in which magnetic entropy change is moderate. An exception is metallic gadolinium that shows the largest magnetic entropy change ($\Delta S_m$ = 9.7 J/kg K for $\mu_0 H$ = 5 T at $T_C$ = 293 K) in any elemental ferromagnets near room temperature due to its high moment ($S$ = 7/2).[1]

The MCE is expected to be much larger in compounds that show a temperature driven first-order paramagnetic to ferromagnetic transition in which $M$ changes discontinuously at $T_C$.[2] In these compounds, an external magnetic field can trigger metamagnetic transition in the paramagnetic state ($T \geq T_C$) which leads to a giant MCE as reported recently in $Gd_5Si_2Ge_2$[3], $MnFeP_{0.45}As_{0.55}$[4], and $Ni_2MnGa$.[5] The colossal magnetoresistive manganites of the formula $R_{1-x}A_xMnO_3$ (R= La, Nd, etc.

and A = Ca, Sr etc.) also joined the race following the reports of a large MCE in $La_{0.75}Sr_{0.25-y}Ca_yMnO_3$ (y = 0.1)[6] and related compounds.[7,8] A comprehensive summary of the MCE in manganites can be found in a recent review by Phan and Yu.[9] Though large lattice heat capacity ($C_p$) of these oxides can lower the adiabatic temperature change ($\Delta T_{ad} = -\frac{T}{C_p}\int_0^H \left(\frac{\partial M}{\partial T}\right) HdH$) manganites have many advantages over other materials: 1. They are relatively cheaper, chemically stable and can be easily prepared by the standard solid state or soft chemistry routes compared to expensive rare earth or inter-metallic based alloys, 2. The $T_C$ can be tuned anywhere between 100 K and 400 K by controlling the hole-doping (x) or the average ionic radii ($\langle r_A \rangle$) of the R and A cations, 3. These materials exhibit either a second-order or a first -order paramagnetic to ferromagnetic transition depending on x and $\langle r_A \rangle$[10] and some compositions exhibit structural transition in the vicinity of ferromagnetic transition that could influence the MCE because of unusual strong coupling between lattice and magnetic degrees of freedom.[11] 4. Besides a second- order paramagnetic to ferromagnetic transitions at high temperature, Sr-based compounds with x = 0.5 such as $Pr_{0.5}Sr_{0.5}MnO_3$ show first-order lattice-coupled ferromagnetic to antiferromagnetic transition at low temperature. Application of an high external magnetic field not only transforms the antiferromagnetic phase into a ferromagnetic phase but also changes structural symmetry and volume which can lead to enhanced magnetic entropy $\Delta S_m$, as shown for $Pr_{0.46}Sr_{0.54}MnO_3$.[12]

In this report, we take an alternative approach to achieve a large MCE by exploiting the nanoscale phase separation found in some narrow band width manganites. We have investigated magnetocaloric effect in $Sm_{1-x}Sr_xMnO_3$ series (x =

0.3, 0.4, and 0.5) whose magnetic phase diagram has been studied in detail.[13,14] The low temperature ground state of $Sm_{1-x}Sr_xMnO_3$ changes from a metallic ferromagnet ($0.35 \le x < 0.5$) to a semiconducting and A-type antiferromagnet for $x \ge 0.5$.[15] In the critical composition x = 0.5 where $T_C = T_N$ ($T_N$-Neel temperature), a long-range charge-orbital ordering also develops below $T_N$.[15] The paramagnetic phase of these Sm based manganites is unusual as suggested by the early work of F. Borges *et al.*[16] who reported a field-induced metamagnetic transition, i.e., a rapid increase of the magnetization above a threshold magnetic field in the paramagnetic state and a large deviation of the inverse susceptibility much above $T_C$. Borges *et al.*[16] suggested the possibility of weakly interacting superparamagnetic clusters for $T >> T_C$. Small Angle Neutron Scattering (SANS) study in $Sm_{0.55}Sr_{0.45}MnO_3$ by J.M. De Teresa *et al.*[17] supported the presence of ferromagnetic nano clusters of about ≈ 0.8 nm in size and stable longer than ~1 ps in zero field. A strong diffusive x-ray scattering below 300 K (> $T_C$ = 118 K) in x = 0.45 was reported earlier and it was attributed to the short-range charge-orbital clusters.[18] Thus, it seems that the paramagnetic state of x = 0.45 has a mixture of short-range charge ordered regions and nanometer size ferromagnetic clusters. The field-driven metamagnetic transition observed in these mateials[19] can lead to a large change in magnetic entropy ($\Delta S_m$). However, MCE in these materials has not been studied in detail. In the course of our investigation of Sm based compounds, we came across the recent work by Sarkar et al.[19] who observed a huge change in magnetic entropy $\Delta S_m$ = 7.9 J/kg K in $\mu_0 H$ = 1 T at $T$ = 120 K in a single crystal of x = 0.52. According to the recently proposed phase diagram, x = 0.52 composition is A-type antiferromagnetic below $T_N$ =120 K.[17] In this report, we focus on the magnetic and magnetocaloric effects in lower compositions, x = 0.3, 0.4, and 0.5.

Polycrystalline samples of $Sm_{1-x}Sr_xMnO_3$ (x = 0.30, 0.40 and 0.50) were prepared by the standard solid state route and characterized by the x-ray diffraction analysis. Magnetic measurements over a wide temperature ($T$ = 10 - 400 K) and magnetic-field ($\mu_0 H$ = 0 - 5 T) range were carried out on the samples using a vibrating sample magnetometer probe in a commercial superconducting cryostat (Physical and Magnetic Property Measurement System, Quantum Design Inc, USA)

Figure 1 shows the temperature dependence of the magnetization ($M(T)$) at $\mu_0 H$ = 0.1 and 5 T for (a) x= 0.3, (b) x = 0.4 and (c) x = 0.5. The $M$-$T$ data at $\mu_0 H$ = 0.1 T suggest that all these compounds undergo a paramagnetic to ferromagnetic transition upon cooling and exhibits a small hysteresis around the transition temperature while warming. The Curie temperature ($T_C$) identified from the inflection point of $M$-$T$ curves while cooling are $T_C$ = 87 K, 118 K and 99 K for x = 0.3, 0.4 and 0.5 respectively. The magnetic transition is rather sharp in case of x = 0.4 compared to the other two compositions. The x = 0.4 also shows a clear peak around $T$ = 30 K << $Tc$ possibly related to increase in the coercivity.[18,20] While the magnetic transition is more broadened under $\mu_0 H$ = 5 T for x = 0.3, the x = 0.4 composition shows a clear upward shift of the $T_C$ to 157 K. The temperature dependence of the inverse susceptibility ($1/\chi$) for all the composition are shown in the respective insets along with the Curie-Weiss fit ($1/\chi = C/(T-\theta_p)$) drawn in red line. The $1/\chi$ clearly deviates from the Curie-Weiss behavior several tens of kelvins above the $T_C$ for all the compositions. The x = 0.4 shows the strongest deviation, which starts around 310 K, far above $T_C$ = 118 K. The observed effective moments $P_{eff}$ = 5.88, 5.8 and 5.22 $\mu_B$ are much larger than expected theoretically ($P_{th}$ = 4.61, 4.51, and 4.44 $\mu_B$ for x = 0.3.

0.4 and 0.5, respectively) assuming $P_{eff}$ ($Sm^{3+}$) = 0.71 $\mu_B$. Borges et al.[19] found a much larger $P_{eff}$ = 19 $\mu_B$ in x = 0.35 when $1/\chi$ vs. $T$ data was extended up to 600 K.

Figure 2 (a) shows the magnetization ($M$) isotherms at different temperatures and Fig. 2(b) shows $H/M$ versus $M^2$ (Arrott plot) for x = 0.3. We have taken magnetization isotherms at temperature interval of $\Delta T$ = 3 K between 130 K and 60 K and at $\Delta T$ = 5 K in other temperature range. The $M$ varies linearly with $H$ above 180 K in the paramagnetic state but shows a Langevin- type superparamagnetic behavior (gradual reorientation of superaparamagnetic moment towards the field direction with increasing field strength) between 180 K and 100 K. Below 80 K, the compounds is a soft ferromagnet with a rapid increase at fields less than 0.1 T followed by a slow approach to saturation at high fields. The saturation magnetization obtained from the extrapolation of the high field magnetization to the origin ($\mu_0 H$ = 0 T) is $M_{sat}$ = 3.5 $\mu_B$ which is closer to the theoretical value of 3.7 $\mu_B$. The Arrot plot shows a positive slope near the Curie temperature suggesting that the transition is second-order according to the Banarjee's criteria.[21] In contrast to x = 0.3, $M(H)$ of x = 0.4 (Fig. 2(c)) shows a metamagnetic behavior- a rapid increase above a critical field ($H_c$) in the paramagnetic state between 170 K and 125 K though it is linear above 180 K. The transition from low- to high magnetic moment state is reversible upon decreasing the field with small hysteresis of width (0.01 T) at 135 K (not shown here). The critical field decreases with lowering temperature, from $\mu_0 H_c$ = 3.9 T at 170 K to $\mu_0 H_c$ = 0.4 T at 127 K. The saturation magnetization at 10 K reaches a value of 3.25 $\mu_B$ which is slightly lower than 3.6 $\mu_B$ expected theoretically. The Arrott plot (Fig. 2(d)) shows a negative slope at temperature above and near $T_c$ which suggests that the field-induced para-ferromagnetic transition is first-order. This satisfies one of the criteria proposed for a material to be a good magnetocaloric material around the Curie temperature.[1]

The *M-H* curves and the Arrott plot for x = 0.5 is shown in Fig. 3 (a) and 3(b), respectively. This sample also shows a metamagnetic behavior in the *M-H* curve between 180 K and 125 K but the transition is rather broad compared to x = 0.4. A ferromagnetic like *M-H* behavior is observed below 90 K. It is to be noted that the extrapolation of high-field data to the origin, gives $M_s$ = 1.88 $\mu_B$ which is far below expected theoretical value, 3.5 $\mu_B$ and this value has not been reached even at the highest field used. The Arrott plot exhibits a negative slope for certain field range above 125 K which suggests the field-induced transition is first order in nature. Since materials with low hysteresis loss is preferred for the application, we compare the hysteresis in *M-H* during increasing and decreasing strength of the magnetic field for all three compositions at *T* = 135 K (Fig. 4(b)). It is seen that the x = 0.4 composition exhibits a small hysteresis but the other two compounds show negligible hysteresis. The width of the hysteresis in x = 0.4 is temperature dependent : it increases with lowering temperature below 160 K down to 120 K and then decreases as shown in Fig. 4(a). The other two compositions did not show hysteresis at all temperatures. The variation in the critical field $H_c$ which corresponds to the onset of metamagnetic transition (determined from the intersection point of the linear increase of *M* at low magnetic fields and the rapid increase at the intermediate fields) with temperature is plotted in inset of Fig. 4(b).

From the magnetization isotherms measured at discrete temperatures, $\Delta S_m$, can be approximated as $\Delta S_M = \frac{1}{\Delta T}\left[\int_0^H M(T+\Delta T, H)\,dH - \int_0^H M(T,H)\,dH\right]$. In order to evaluate the magnetic entropy from our isothermal *M(H)* curves, we took a numerical approximation to the above integral. The method is to use the isothermal magnetization measurements at small discrete field intervals, $|\Delta S_m| = \sum_i \frac{M_i - M_{i+1}}{T_{i+1} - T_i}\Delta H_i$ where, $M_i$ and $M_{i+1}$ are the experimental values of the magnetization at temperatures, $T_i$ and $T_{i+1}$, respectively, under an applied magnetic field $H_i$. Using the above equation, the entropy change associated with the magnetic field variation calculated from the *M-H* curve at various fixed temperatures is plotted in Fig.5 (a), 5(b) and 5(c) for the three compositions x = 0.3, 0.4 and 0.5 respectively. For x = 0.3, $\Delta S_m$ at $\mu_0 H$ = 1 T shows a broad maximum around $T_C$ = 87 K with a maximum value of $\Delta S_m$ = 0.85 J/kg K and the maximum shift to 97 K at 3 T. Though the magnitude of $\Delta S_m$ increases with *H*, the temperature corresponding to the maximum remains nearly the same at $\mu_0 H$ = 5 T. The $\Delta S_m$ of x = 0.4 (Fig. 5(b)) exhibits a rather prominent peak compared to other two compounds. The peak shifts from 117 K at $\mu_0 H$ = 1 T to 127 K at $\mu_0 H$ = 5 T. Interestingly, this composition shows the largest entropy change $\Delta S_m$ = 6.2 J/kg K at $\mu_0 H$ = 5 T and $T$ = 127 K followed by $\Delta S_m$ = 3.3 J/kg K at 97 K for x = 0.3 (Fig. 5(a)) and $\Delta S_m$ = 2.3 J/kg K at 125 K for x = 0.5 (Fig. 5(c)). The $\Delta S_m$ of x = 0.4, unlike the other two compounds, exhibits a sharp drop just below the peak values for each magnetic field and it weakly depends on the field for $T$ < 50 K. The x = 0.3 and 0.5 samples show a field dependent entropy over a wide temperature range below the maximum compared to x= 0.4. The relative cooling power (RCP), a measure of heat transferred by a magnetic refrigerant, is obtained by the product of the peak value of the entropy ($\Delta S_m^{peak}$) and the full width at half maximum ($\delta T_{FWHM,}$), i.e., $RCP(T,\Delta H) = \Delta S_m\ (T,\Delta H)\ \delta T_{FWHM}$ . The larger the $\delta T_{FWHM}$, the better is the cooling-capacity. The *RCP* of x = 0.3 (-326.3J/Kg) is higher than x = 0.4 (- 269.5 J/Kg) and x = 0.5 (-231 J/Kg) at 5 T.

The largest $\Delta S_m$ observed in x = 0.4 is clearly related to the field-driven metamagnetic transition occurring in the paramagnetic state. The large deviation of the inverse susceptibility $1/\chi$ much above $T_C$ (= 118 K) found in this compound suggests existence of superparamagnetic clusters in the paramagnetic state. Our detailed analysis of *M-H* data at different temperatures above $T_C$ cannot be fitted $M/M_s$ vs. $\mu B/k_BT$ as expected for non-interacting superparamagnetic clusters but to $M/M_s$ vs. $\mu B/k_B(T-\theta)$ where $\theta$ is measure of interaction between clusters. The detailed analysis of magnetization data is beyond the scope of this letter and will be reported later. We note that small angle neutron scattering study in x = 0.45 indeed suggested ferromagnetic clusters of $\approx$ 0.8 nm size imbedded in short-range charge/orbital ordered paramagnetic matrix. These nanometer size clusters are weakly interacting in low magnetic fields but as the applied magnetic field increases, interaction between ferromagnetic nano clusters increases. Above a critical field ($H_C$), the size of the clusters increase abruptly and eventually they coalesce at higher fields. Thus, microscopically inhomogeneous paramagnetic state transforms into homogeneous ferromagnetic phase at higher fields. This process leads to the rapid increase of the magnetization above $H_C$ as observed in the *M* versus *H* data for x = 0.4 that results in the largest change in the magnetic entropy. However, the observed value of $\Delta S_m$ = 2 J/kg K at $\mu_0H$ = 1 T is a factor of two smaller than the value reported in single crystalline[21] $Sm_{0.52}Sr_{0.48}MnO_3$ and it is possibly due to smearing of the metamagnetic transition by disorders such as grain boundaries in polycrystalline sample. The *M-H* curve in our polycrystalline compounds exhibits much smaller hysteresis which is preferred for magnetic refrigeration compared to single crystals that show a sharp metamagnetic transition but with pronounced hysteresis.[17] The magnetization of x = 0.5 even at 5 T is 2.5 $\mu_B$ is lower than the value (= 3.5 $\mu_B$ / f.u) expected for complete parallel alignment of Mn spins. It suggests the low temperature ground state of x = 0.5 is different from that of x = 0.4 and it is possibly inhomogeneous mixture of ferromagnetic and antiferromagnetic clusters. Indeed, single crystalline study shows that the low temperature ground state changes from ferromagnetic to A-type antiferromagnetic at x = 0.5 and in this critical composition both $T_C$ and $T_N$ may coincide and long range charge-ordering also develops below 80 K.[16] The decrease in the field-induced magnetic entropy in this composition can be attributed to the existence of stronger charge-orbital correlation above $T_C$ and smaller size of the ferromagnetic clusters.

In summary, we have studied magnetic and magnetocaloric properties in $Sm_{1-x}Sr_xMnO_3$ (x = 0.3, 0.4, and 0.5) and found the largest magnetocaloric effect in x = 0.4 among the three compositions. The largest MCE observed in x = 0.4 is related to the existence of ferromagnetic nano clusters in the paramagnetic state that grow and coalesce with increasing strength of the external magnetic field in contrast to the case x = 0.3 which shows a gradual reorientation of superparamagnetic moment towards the field direction. Our results suggest that other manganites that show a field-induced metamagnetic transition in the paramagnetic state such as Cr or Ni doped $Pr_{0.5}Ca_{0.5}MnO_3$[22] or lower band width manganites such as $Eu_{1-x}Sr_xMnO_3$[23] may be attractive candidates to look for a large magnetocaloric effect.

**Acknowledgment**

This work was supported by the NUS-Young Investigator Award (WBS no. R144-000-197-123) to the second author.

**References**

[1] K. A. Gschneidner, Jr., V. K. Pecharsky, and A. O. Tsokol, Rep. Prog. Phys. **68**, 1479 (2005), and references therein.

[2] D. S. Rodbell and C. P. Bean, J. Appl. Phys. **33**, 1037 (1962).

[3] V. K. Pecharsky, and K. A. Gschneidner, Jr. Phys. Rev. Lett. **78**, 4494 (1997); L. Morellon, J. Blasco, P. A. Algarabel, and M. R. Ibarra, Phys.Rev. B **62**, 1022 (2000).

[4] O. Tegus, E. Brück, K. H. J. Buschow, and F. R. de Boer, Nature **415**, 150 (2002).

[5] F. Hu, B. Shen, and J. Sun, Appl. Phys. Lett.**76**, 3460 (2000); F. Casanova, X. Batlle, A. Labarta, J. Marcos, L. Mañosa, and A. Planes. Phys. Rev. B **66**, 100 401 (R) (2002).

[6] Z. B. Guo, Y. W. Du, J. S. Zhu, H. Huang, W. P. Ding, and D. Feng, Phys. Rev. Lett. **78**, 1142 (1997).

[7] X. Bohigas, J. Tejada, E. del Barco, X. X. Zhang, and M. Sales, Appl. Phys. Lett. **73**, 390 (1998); Y. Sun, X. Xu, and Y. Zhang, J. Magn. Magn. Mater. **219**, 183 (2000).

[8] H. Terashita, J. J. Garbe, and J. J. Neumeier, Phys. Rev. B **70**, 094403 (2004); R. Venkatesh, M. Pattabiraman, S. Angappane, G. Rangarajan, K. Sethupathi, J. Karatha, M. F. Morariu, R. M. Ghadimi, and G. Guntherodt, Phys. Rev. B **75**, 224415 (2007); J. S. Amaral, N. J. O. Silva and V. S. Amaral, Appl. Phys. Lett. **91**, 172503 (2007).

[9] M. H. Phan, S. C. Yu, J. Magn. Magn. Mater. **308**, 325 (2007) and references therein.

[10] J. Mira, J. Rivas, F. Rivadulla, C. Vázquez-Vázquez, and M. A. López-Quintela, Phys. Rev. B **60**, 2998 (1999).

[11] H. Terashita, B. Myer, and J. J. Neumeier, Phys. Rev. B **72**, 132415 (2005).

[12] R. Mahendiran, B. G. Ueland, P. Schiffer, A. Maignan, C. Martin, M. Hervieu, B. Raveau, M. R. Ibarra, and L. Morellon, www. arXiv:cond-mat/0306223 (2003).

[13] C. Martin, A. Maignan, M. Hervieu, and B. Raveau, Phys. Rev. B **60**, 12191 (1999); A. I. Kurbakov, A. V. Lazuta, V. A. Trounov, I. I. Larionov, C. Martin, A. Maignan, and M. Hervieu, Phy. Rev. B. **72**, 184432 (2005).

[14] L. M. Fisher, A. V. Kalinov, I. F. Voloshin, N. A. Babushkina, D. I. Khomskii, Y. Zhang, and T. T. M. Palstra, Phy. Rev. B. **70**, 212411 (2004); M. Egilmez, K. H. Chow, J. Jung, and Z. Salman, Appl. Phys. Lett. **90**, 162508 (2007).

[15] Y. Tomioka, H. Hiraka, Y. Endoh, and Y. Tokura, Phys. Rev. B **74**, 104420 (2006).

[16] R. P. Borges, F. Ott, R. M. Thomas, V. Skumryev, J. M. D. Coey, J. I. Arnaudas, L. Ranno, Phy. Rev. B. **60**, 12847 (1999).

[17] J. M. De Teresa, M. R. Ibarra, P. Algarabel, L. Morellon, B. Garcia-Landa, C. Marquina, C. Ritter, A. Maignan, C. Martin, B. Raveau, A. Kurbakov, and V. Trounov, Phys. Rev. B **65**, R100403 (2002).

[18] Y. Tokura and N. Nagaosa, Science **288**, 462 (2000); E. Saitoh, Y. Tomioka, T. Kimura, and Y. Tokura, J. Phys. Soc. Jpn. **69**, 2403 (2000).

[19] P. Sarkar, P. Mandal, and P. Choudhury, Appl. Phy. Lett. **92**, 182506 (2008).

[20] E. M. Levin and P. M. Shand, J. Magn. Magn. Mater. **311**, 675 (2007).

[21] S. K. Banerjee, Phys. Lett. **12,** 16 (1964).

[22] R. Mahendiran, M. Hervieu, A. Maignan, C. Martin, and B. Raveau, Solid State Commun. **114**, 429 (2000).

[23] L. Jia, G. J. Liu, and J. Z. Wang, J. R. Sun, H. W. Zhang, and B. G. Shen, Appl. Phys. Lett. **89**, 122515 (2006).

## Figure captions

**Fig. 1** Temperature dependence of the magnetization of the $Sm_{1-x}Sr_xMnO_3$ (a) x = 0.3, (b) x = 0.4 and (c) x = 0.5 at magnetic field of $\mu_0 H$ = 0.1 Tand 5 T.

**Fig. 2 (a**) *M-H* isotherms and (b) Arrott plot for the composition x = 0.3. The bottom panels show the respective plots for x = 0.4. *M-H* isotherms were taken at a temperature interval of $\Delta T$ = 3 K interval in between 130 K and 90 K, i.e. in the regime of magnetic phase transition and at $\Delta T$ = 5 K at other temperatures.

**Fig. 3** (a) *M-H* isotherms and (b) Arrott plot for the composition x = 0.5. The data were taken at $\Delta T$ = 3 K interval in between 130 K and 100 K and at $\Delta T$ = 5 K interval away from the phase transition**.**

**Fig. 4 (a)** *M-H* isotherms for the composition x = 0.4 and (b) comparison of *M-H* curve at 135 K for x = 0.3, 0.4 and 0.5. The inset of Fig. 4(b) shows the variation of $H_c$ with temperature. The low field paramagnetic (PM) is state is an inhomogeneous mixture of non size ferromagnetic and charge ordered clusters and the high field state has a larger fraction of ferromagnetic phase.

**Fig. 5** The temperature dependence of the change in the magnetic entropy $\Delta S_m$ at different magnetic fields for x = 0.3 (a), 0.4 (b) and 0.5 (c), respectively.

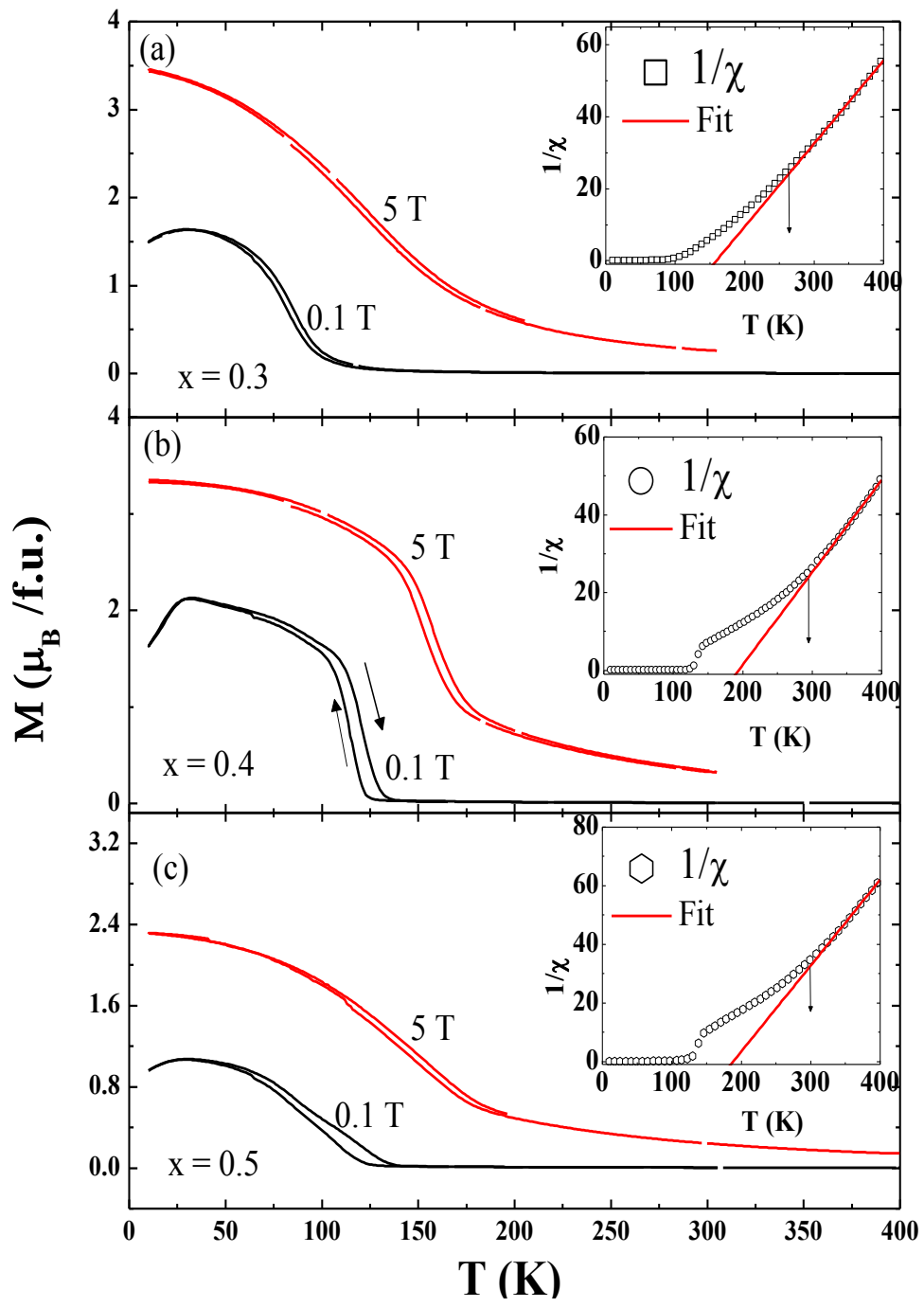

Fig. 1

A. Rebello et al.

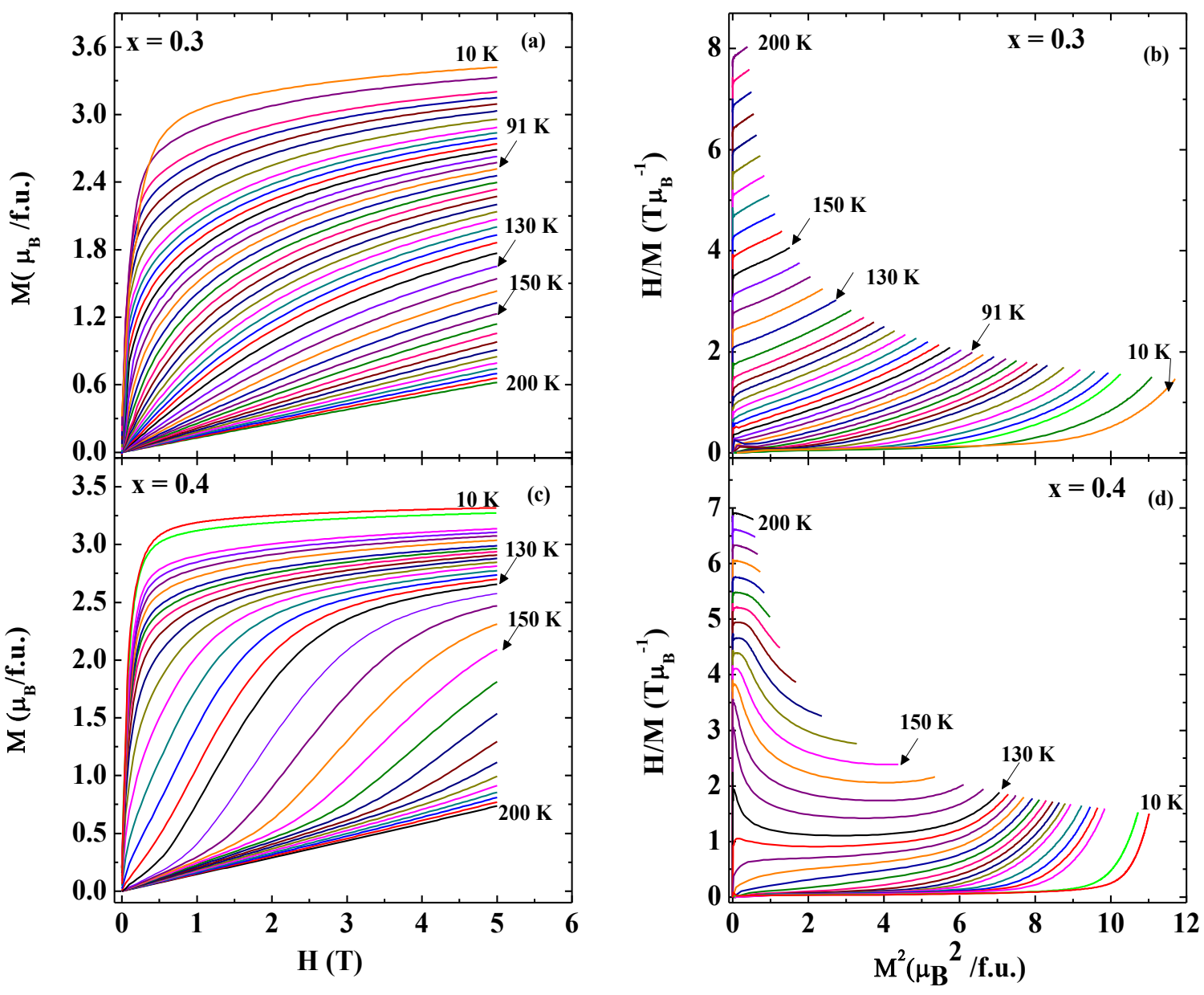

Fig. 2

A. Rebello et al.

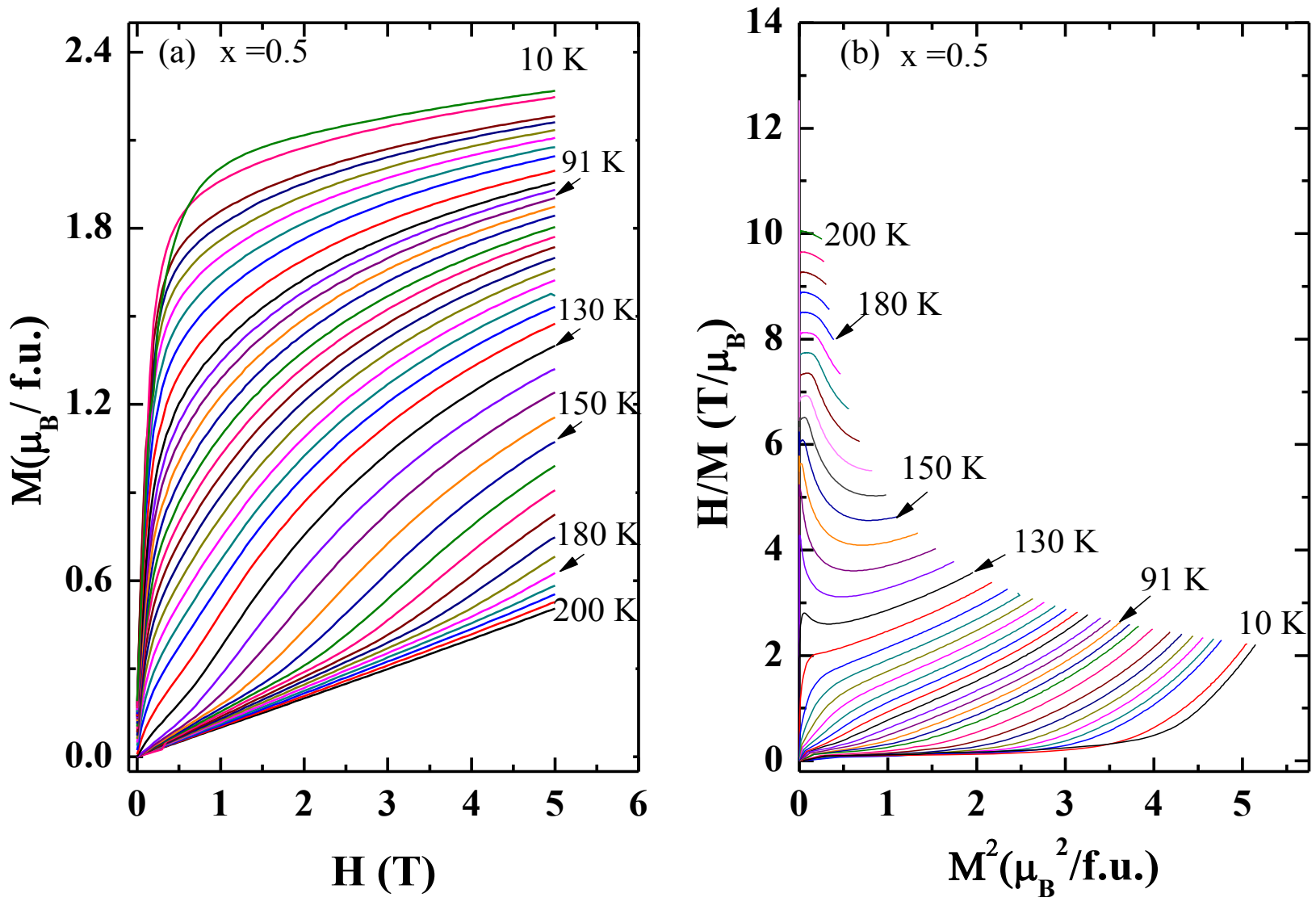

Fig. 3

A. Rebello et al.

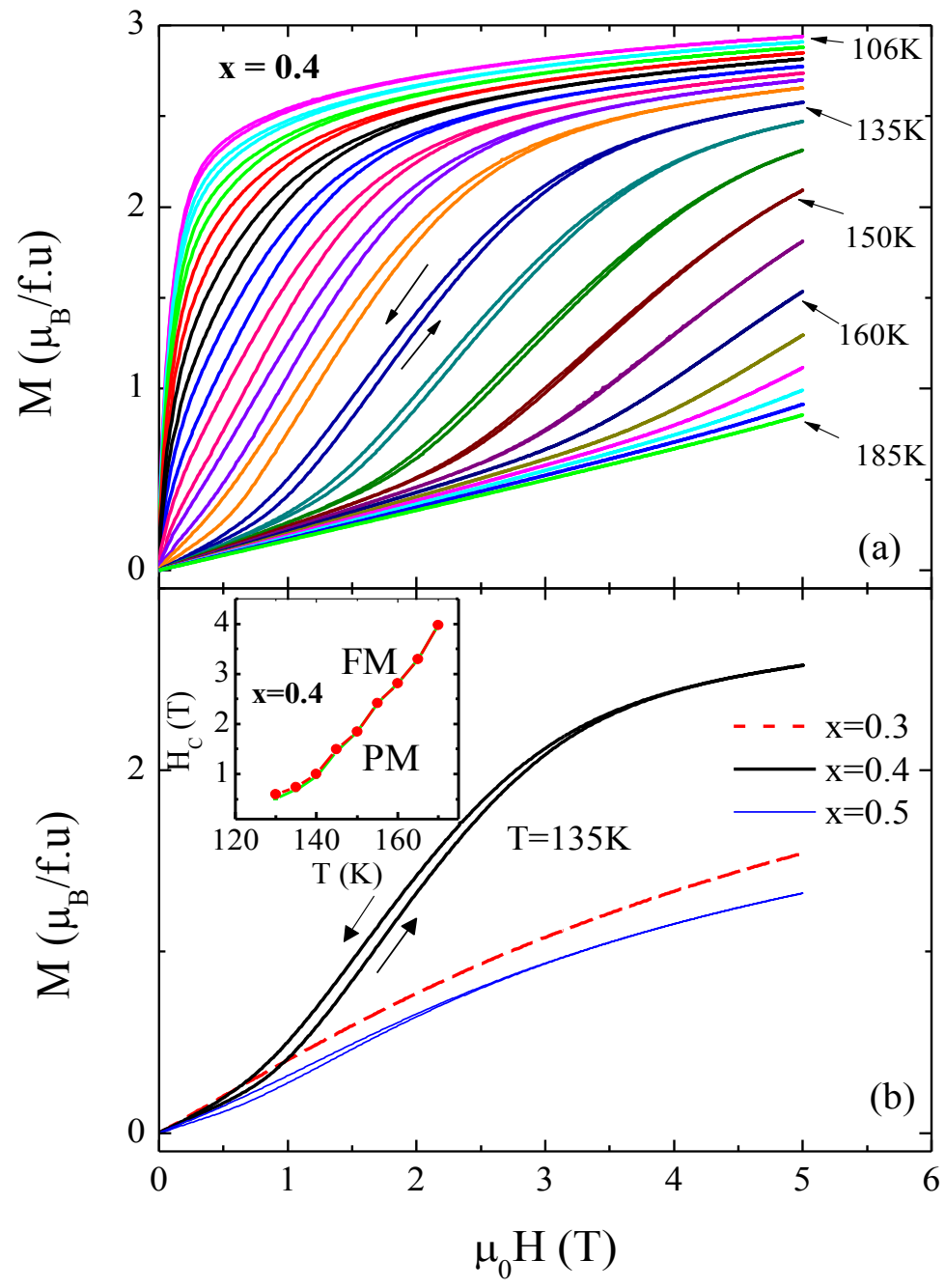

Fig. 4

A. Rebello et al.

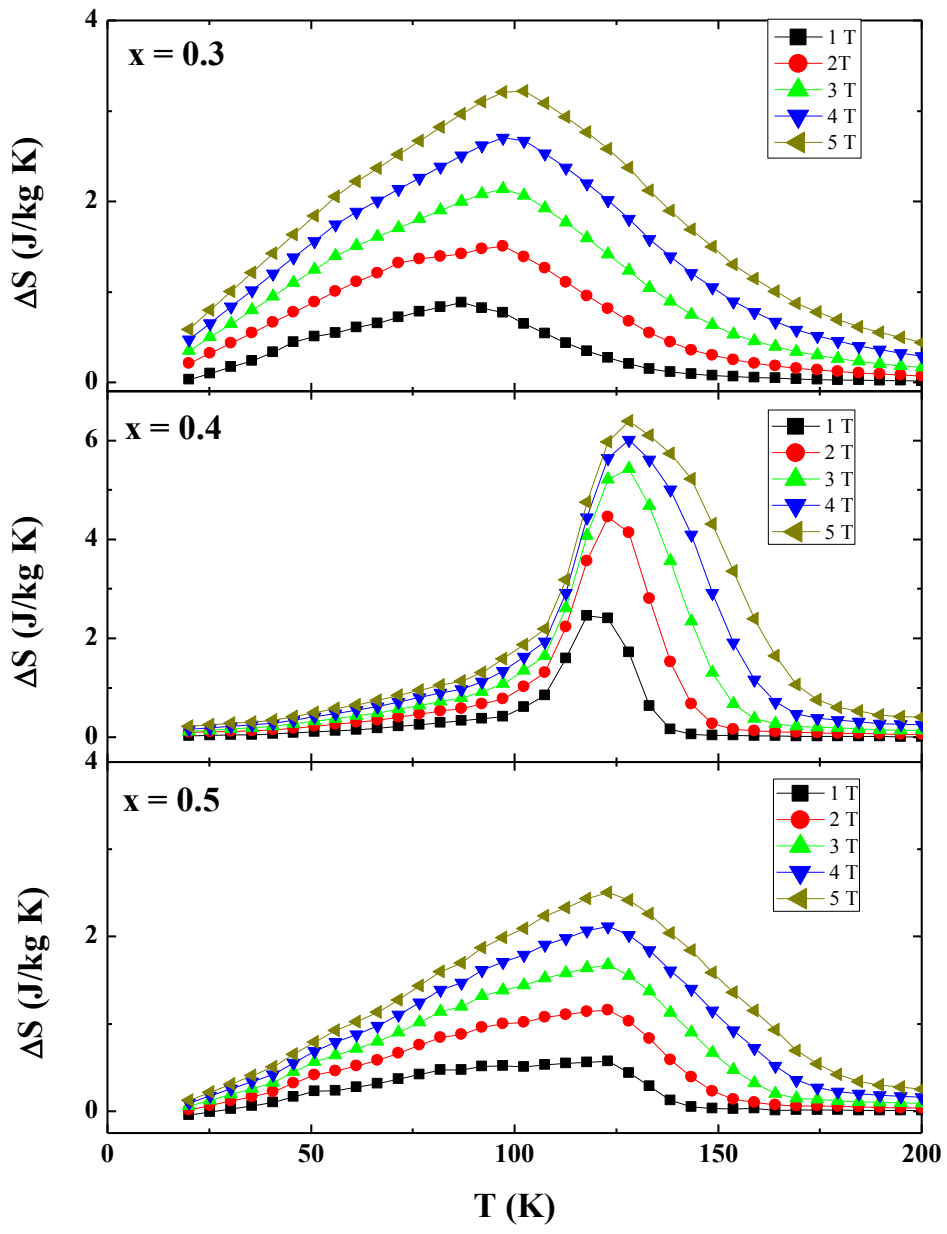

Fig. 5

A. Rebello et al.